\documentclass[twocolumn,preprintnumbers,amsmath,amssymb,showpacs]{revtex4}

\usepackage{amssymb}
\usepackage{graphicx}
\usepackage{dcolumn}
\usepackage{bm}
\usepackage{SIunits}
\usepackage{amsmath}

\begin{document}

\title{Charge-SQUID and Tunable Phase-slip Flux Qubit}

\author{Hu Zhao$^{1,2}$}
\email{zhaohu09@mails.tsinghua.edu.cn}
\author{Tie-Fu Li$^{1,2,3}$}
\author{Jian She Liu$^{1,2}$}
\author{Wei Chen$^{1,2}$}
\email{weichen@mail.tsinghua.edu.cn}
\affiliation{$^{1}$Institute of Microelectronics, Tsinghua University, Beijing 100084, China}
\affiliation{$^{2}$Tsinghua National Laboratory of Information Science and Technology, Beijing 100084, China}
\author{J. Q. You$^{3}$}
\email{jqyou@csrc.ac.cn}
\affiliation{$^{3}$Beijing Computational Science Research Center, Beijing 100084, China}
\date{\today}

\begin{abstract}
\noindent A phase-slip flux qubit, exactly dual to a charge qubit, is composed of
a superconducting loop interrupted by a phase-slip junction. Here we propose a tunable phase-slip flux qubit by replacing the phase-slip junction with a charge-related superconducting quantum interference device (SQUID) consisting of two phase-slip junctions connected in series with a superconducting island. This charge-SQUID acts as an effective phase-slip junction controlled by the applied gate voltage and can be used to tune the energy-level splitting of the qubit. Also, we show that a large inductance inserted in the loop can reduce the inductance energy and consequently suppress the dominating flux noise of the phase-slip flux qubit. This enhanced phase-slip flux qubit is exactly dual to a transmon qubit.
\end{abstract}
\pacs{   }
\maketitle

\begin{figure}[h]\center
　　\includegraphics[width=7.6cm]{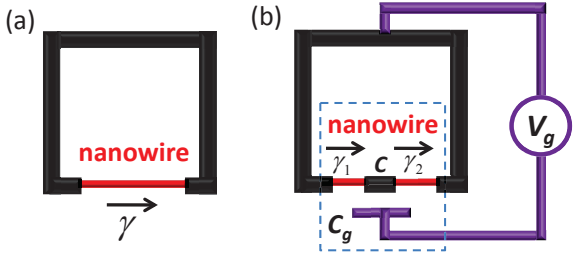}\\
　　\caption{(a)~A phase-slip flux qubit, where a superconducting loop is interrupted by a nanowire acting as a phase-slip junction. Similar to a Josephson junction, this junction also has a phase drop $\gamma$. (b)~A tunable phase-slip flux qubit, where a central superconducting island is connected by two phase-slip junctions, so as to form a loop with a segment of superconductor wire. Also, this island is biased by a voltage $V_g$ via a gate capacitance $C_g$. In both (a) and (b), an external magnetic flux is also applied to the qubit loop.}
\label{fig1}
\end{figure}

\indent Using superconducting qubits to make a quantum computer
is considered as one of the most promising methods for achieving scalable quantum
computation~(see, e.g., \cite{You01,Clarke01}). The basic element in a superconducting qubit is
a device called Josephson junction~\cite{Josephson01}, which is formed by two superconductors
connected with a thin insulator layer. According to the ratio of $E_J/E_C$, where $E_J$
is the Josephson energy of the device and $E_C$ is its charging energy, different kinds of
superconducting qubits were proposed, including charge
qubit~\cite{Nakamura01}, flux qubit~\cite{Mooij01,Friedman}, phase qubit~\cite{Martinis01,Yu},
and other enhanced superconducting qubits such as the charge-flux qubit~\cite{Vion}, gradiometer-based flux qubit~\cite{DiVincenzo}, low-decoherence flux qubit~\cite{You02, Steffen}, transmon qubit~\cite{transmon} and fluxonium qubit~\cite{fluxonium}. Phase-slip
phenomenon has been studied extensively around the critical temperature
of superconductors, where it is dominated by thermal activations~{\cite{Tinkham01,Tinkham02}}.
However, as the temperature drops, thermal activation is
suppressed and quantum phase-slip tunneling will happen. Based
on the quantum phase-slip tunneling, a new device called phase-slip
junction, which is exactly dual to Josephson junction, was proposed~\cite{Mooij02}.
A superconducting loop containing a phase-slip junction makes a new type of superconducting
qubit called the phase-slip flux qubit~\cite{Mooij02,Mooij03}; see also Fig.~\ref{fig1}(a). As a possible experimental realization, it was proposed in \cite{Mooij03} to produce a phase-slip flux qubit using the material SiNbx.
Because the energy-level splitting of a phase-slip flux qubit is sensitive to the
coherence length $\xi$ of a superconductor and the resistance per unit length $R_n$ (k$\Omega$/nm),
the phase-slip flux qubit is hard to fabricate and measure.
Moreover, the phase slip is the fluctuation of the phase, so strong
fluctuations require materials with a higher degree of disorder. This disorder
can cause electron localization, which may suppress
superconductivity of the material. Therefore, a suitable material must be found, which has the maximum disordering
but remains superconducting. Fortunately, two materials have been found to fit the above
criterion, i.e., TiN~\cite{Sacepe01} and InO$_{\rm x}$~\cite{Sacepe02}. Recently, a phase-slip flux
qubit was produced and coherent phase-slip
tunneling was also observed using the material InO$_{\rm x}$~\cite{Oleg01}.

In this paper, a tunable phase-slip flux qubit based on charge-SQUID (here SQUID is abbreviated from superconducting quantum interference device)
[blue dashed box in Fig.~\ref{fig1}(b)] is proposed,
which is formed by two phase-slip junctions connected in series with a superconducting island biased by a gate voltage.
This device is similar to a single-charge transistor~\cite{Zorin01}.
By changing the gate voltage of the charge-SQUID, the energy-level
splitting of the phase-slip flux qubit can be tuned. Thus, our controls on the property of
this tunable phase-slip flux qubit can have two degrees of freedom, i.e., the magnetic flux threading the loop and the gate voltage applied on the island.
Also, if we increase the loop inductance, the energy level of the phase-slip flux qubit will become
more flatten, and then the qubit becomes less sensitive to the flux noise.
This enhanced phase-slip flux qubit can reduce the effect of flux noise,
which is dual to a transmon qubit.

The relation between the voltage and charge in a phase-slip
junction can be written as $V=V_c\sin(2\pi q)$, where $V_C=2\pi E_S/2e$ is
the critical voltage of the phase-slip junction, with $E_S$ being the phase-slip quantum tunneling rate, and $q$ is the number of Cooper pairs tunneling through the phase-slip junction. A weak link
in a superconducting loop can cause flux tunneling. If one puts a phase-slip junction into a superconducting loop,
it forms a phase-slip flux qubit~\cite{Mooij02,Mooij03}; see also Fig.~\ref{fig1}(a). Below we first derive the Hamiltonian of the phase-slip flux
qubit given in \cite{Mooij02,Mooij03}, and then show how to achieve a tunable phase-slip flux qubit by introducing a charge-SQUID [see Fig.~\ref{fig1}(b)].

For the phase-slip flux qubit in Fig.~\ref{fig1}(a), the kinetic energy is the flux energy stored in the loop:
$T=\frac{1}{2}(L_k+L_g)I^2$, where $L_k$ is the kinetic inductance of the phase-slip junction, $L_g$ is
the geometric inductance of the loop, and $I$ is the supercurrent through the phase-slip junction,
which is also the persistent current in the loop. A Josephson junction has the phase-voltage ralation
$V=\hbar\dot{\phi}/2e$, while a phase-slip junction has the charge-current relation $I=2e\dot{q}$. To obtain
the Lagrangian of the circuit, we also need the potential energy. Similar to the Josephson coupling energy
$E_J(1-\cos\phi)$, the potential energy of the phase-slip junction is $U=\int_{0}^{t}IVdt=E_S[1-\cos(2\pi q)]$.
Excluding the constant term, one can write the potential energy as $U=-E_S\cos(2\pi q)$.

\indent The Lagrangian of the circuit is $L=T-U=\frac{1}{2}(L_k+L_g)(2e\dot{q})^2+E_S\cos(2\pi q)$, and
the canonical momentum is $p=\partial L/\partial\dot{q}=(L_k+L_g)(2e)^2\dot{q}=(L_k+L_g)2eI$.
Then, the Hamiltonian of the circuit is given by
\begin{gather}
H=p\dot{q}-L=\frac{1}{2}(L_k+L_g)I^2-E_S\cos(2\pi q),
\end{gather}
where the momentum $p=(L_k+L_g)2eI$ is canonically conjugate to the Cooper-pair number $q$.

Assume that the phase drop across the phase-slip junction is $\gamma$. It is related to the kinetic
inductance of the phase-slip junction by $L_kI=(\Phi_0/2\pi)\gamma$, where $\Phi_0=h/2e$ is the flux quantum.
Similarly, the phase drop $\gamma'$ related to the geometric inductance of the loop is given by
$L_gI=(\Phi_0/2\pi)\gamma'$. Thus, $(L_k+L_g)I=(\Phi_0/2\pi)(\gamma+\gamma')$, namely,
\begin{gather}\label{E1}
I=\frac{(\Phi_0/2\pi)}{L_k+L_g}(\gamma+\gamma').
\end{gather}
For the superconducting loop interrupted by a phase-slip junction, the fluxoid quantization condition yields
\begin{gather}\label{E2}
\gamma+\gamma'+2\pi f=2\pi n,
\end{gather}
where $n$ denotes an integer corresponding to a fluxoid state, and $f=\Phi_{\rm ext}/\Phi_0$, with $\Phi_{\rm ext}$ being the external flux applied to the loop. From Eqs.~(\ref{E1}) and (\ref{E2}),
it follows that
\begin{gather}
H=E_L(n-f)^2-E_S\cos(2\pi q),
\end{gather}
where $E_L=\Phi_0^2/2(L_k+L_g)$ is the inductance energy of the superconducting loop with a phase-slip junction.
In the conventional phase-slip flux qubit, the kinetic inductance $L_k$ of the phase-slip junction is much larger than
the geometric inductance $L_g$ of the loop, so the inductance energy can be approximated as \cite{Mooij03} $E_L=\Phi_0^2/2L_k.$
Representing $\cos(2\pi q)$ in the fluxoid basis states, one can then write the Hamiltonian as \cite{Mooij02}
\begin{gather}
H=E_L(n-f)^2-\frac{E_S}{2}\sum_{n}(|n\rangle\langle n+1|+|n+1\rangle\langle n|).\
\end{gather}
It is clear that $E_S$ represents the flux tunneling rate of the phase-slip junction.

For the charge-SQUID, i.e., the part surrounded by a blue dashed rectangle in Fig.~\ref{fig1}(b), a central
superconducting island is connected
by two phase-slip junctions, similar to the single-charge transistor proposed in \cite{Zorin01}. Note that these two phase-slip junctions connect
the source and drain in \cite{Zorin01}. This is different from our proposed tunable
phase-slip flux qubit where we connect these two junctions with a segment of superconducting wire to form a superconducting loop. Moreover, the regime with both sufficiently small gate capacitance $C_g$ and island self-capacitance
$C$ was considered in \cite{Zorin01}, so as to demonstrate the exact duality between its device and the dc-SQUID. Instead,
our circuit is designed to have a large self-capacitance $C$ for the central island, so that the charging energy of the island
can be neglected.

Here we assume that the two phase-slip junctions in Fig.~\ref{fig1}(b) are identical to each other, so the critical voltages and
kinetic inductance satisfy $V_{C1}=V_{C2}=V_C$ and $L_{k1}=L_{k2}=L_{k}$ for them.
The Hamiltonian of the tunable phase-slip flux qubit can be written as
\begin{gather}\label{E6}
H=\frac{1}{2}(2L_k+L_g)I^2-E_S[\cos(2\pi q_1)+\cos(2\pi q_2)],\
\end{gather}
where $E_S=2eV_C/2\pi$ is the flux tunneling rate of each junction, and $q_i$, with $i=1$ or $2$, is the number of Cooper pairs
tunneling through the $i$th phase-slip junction. Similar to Eq.~(\ref{E2}), the supercurrent is given by
\begin{gather}\label{E7}
I=\frac{(\Phi_0/2\pi)}{2L_k+L_g}(\gamma_1+\gamma_2+\gamma'),\
\end{gather}
where $\gamma_i$, with $i=1$ or $2$, is the phase drop across the $i$th phase-slip junction. Now, the fluxoid quantization condition
becomes
\begin{gather}
\gamma_1+\gamma_2+\gamma'+2\pi f=2\pi n,\
\end{gather}
and then the suppercurrent in Eq.(\ref{E7}) becomes
\begin{gather}\label{E9}
I=\frac{\Phi_0}{2L_k+L_g}(n-f).\
\end{gather}
Moreover, charge conservation on the central island requires that
\begin{gather}
q_1-q_2=C_gV_g/2e,\
\end{gather}
where $V_g$ is the gate voltage applied to the island via the gate capacitance $C_g$. Thus, the total potential energy
$U=-E_S[\cos(2\pi q_1)+\cos(2\pi q_2)]$ can be expressed as
\begin{gather}\label{E11}
U=-2E_S\cos\left(\frac{\pi C_gV_g}{2e}\right)\cos(2\pi q),\
\end{gather}
where $q=(q_1+q_2)/2$. Analogous to a dc-SQUID, these two phase-slip junctions connected to a voltage-biased island can  behave as an effective junction, and the
effective flux tunneling rate is tuned by the gate voltage: $E_S(V_g)=2E_S\cos(\pi C_gV_g/2e)$. Because the
effective phase-slip junction is modulated by the gate-voltage-induced charge, we call this device a charge-SQUID.

Finally, by substituting Eqs.~(\ref{E9}) and (\ref{E11}) into Eq.~(\ref{E6}), the Hamiltonian can be rewritten as
\begin{eqnarray}\label{E12}
H & = & E_L(n-f)^2-E_S(V_g)\cos(2\pi q)\\
  & = & E_L(n-f)^2-\frac{E_S(V_g)}{2}\sum_{n}(|n\rangle\langle n+1|+|n+1\rangle\langle n|),\nonumber
\end{eqnarray}
where $E_L=\Phi_0^2/2(2L_k+L_g)$ is the inductance energy of the superconducting loop with two identical phase-slip
junctions. From the Hamiltonian, we know that the energy-level splitting $E_S(V_g)$ of the phase-slip flux qubit can be tuned
by the gate voltage $V_g$. Theoretically, $E_S(V_g)$ can be tuned from
zero to its maximum value $2eV_C/\pi$.

\begin{figure}[h]\center
　　\includegraphics[width=7.6cm]{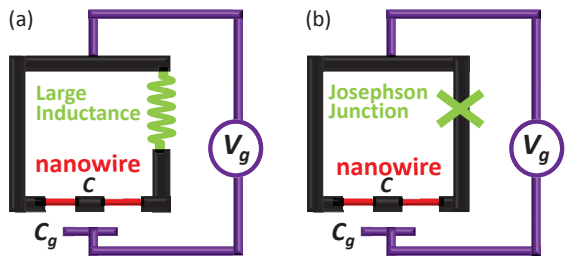}\\
　　\caption{(a)~An enhanced tunable phase-slip flux qubit, where a large inductance is inserted in the loop to reduce the inductance energy of the qubit. (b)~Another possible enhanced tunable phase-slip flux qubit, where the large inductance in (a) is replaced by a Josephson junction which can act as a nonlinear inductance.}
\label{fig2}
\end{figure}

The tunable phase-slip flux qubit is exactly dual to a tunable charge qubit. The latter has
two controlled degrees of freedom: one is the magnetic flux biased to the dc-SQUID for changing
the energy-level splitting and another is the applied gate voltage for shifting the working point.
On the contrary, in our tunable phase-slip flux qubit, the flux bias is used for
shifting the working point, and the gate voltage is for tuning the energy-level splitting.
It is interesting to note that an exchange between two controlled degrees of freedom can yield an exactly dual qubit. Physically, this is
originated from the commutation relation between the phase and the Cooper-pair (charge) number in superconductors.
Similar to the enhanced superconducting qubit in \cite{You02,Steffen}, a large capacitance is shunted to the Josephson junction in a transmon qubit~\cite{transmon}, so as to suppress the main decoherence due to the charge noise. Because of the exact duality between a charge qubit and a phase-slip flux qubit,
if we insert a large inductance into the loop of the tunable phase-slip flux qubit, an enhanced
phase-slip flux qubit can be achieved, where the effects of dominating flux noise in the previous phase-slip flux qubit are reduced.

\begin{figure}[h]\center

　　\includegraphics[width=8.2cm]{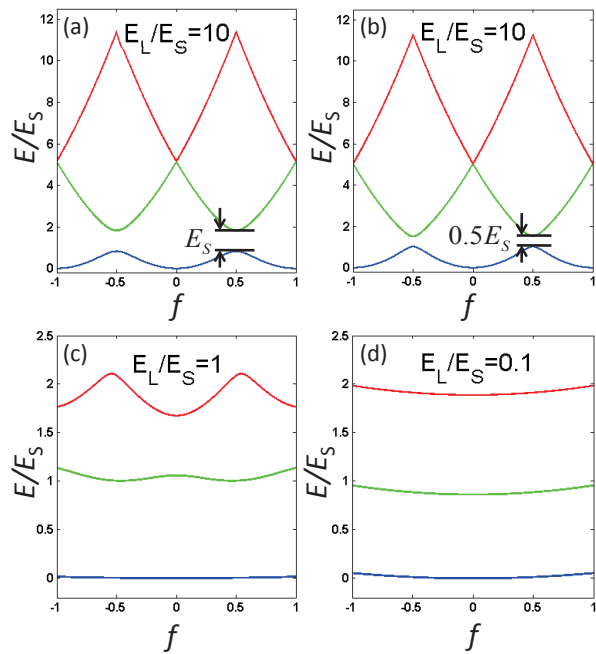}\\
　　\caption{Lowest three energy levels of the phase-slip flux qubit versus the reduced magnetic flux $f$ for
different values of $E_L/E_S$. Here all energies are given in units of $E_S$
and the lowest energy point in the figure is set to be zero.
(a)~Energy levels of a phase-slip flux qubit with $E_L/E_S=10$.
(b)~Energy levels of a tunable phase-slip flux qubit with $E_L/E_S=10$, where
the level splitting at each degeneracy point is tuned to be $0.5E_S$ by changing the gate voltage.
(c)~Energy levels of an enhanced phase-slip flux qubit with $E_L/E_S=1$.
(d)~Energy levels of an enhanced phase-slip flux qubit with $E_L/E_S=0.1$.
From (c) and (d), it is clear that the energy levels become flattened when decreasing the value of $E_L/E_S$.}
\label{fig3}
\end{figure}

According to Eq.~(\ref{E12}), the first term $E_L(n-f)^2$ determines the parabola energy
outlines and the second term $\frac{1}{2}E_S(V_g)\sum_{n}(|n\rangle\langle n+1|+|n+1\rangle\langle n|)$ determines the energy-level splitting. As shown in Fig.~\ref{fig3}(a)-\ref{fig3}(d), if we reduce the inductance energy $E_L$ by increasing the loop inductance [Fig.~\ref{fig2}(a)], energy levels of the enhanced
phase-slip flux qubit will become more flattened. A similar effect may also be achieved by inserting a large Josephson junction in the loop [Fig.~\ref{fig2}(b)] because a Josephson junction can behave as a nonlinear inductance (see, e.g., \cite{You01,Clarke01}). In fact, the larger is the
loop inductance, the more flattened the energy levels will become, but the anharmonicity of the system
will decrease. In other words, the flattened energy levels in the enhanced
phase-slip flux qubit are achieved by sacrificing the anharmonicity
of the system, similar to a transmon~\cite{transmon}. For example, if we set $E_S$ to be 10GHz and consider the intermediate case with $E_S=E_L$, i.e., Fig.~\ref{fig3}(c), we have $(L_k+L_g)\approx300$nH. Actually, the kinetic inductance of a phase-slip junction that exhibits quantum phase slip is of the order $\sim1$nH~\cite{Mooij03}, so the geometric inductance $L_g$ should be in the order of 100nH.

It is known that a superconducting qubit is a macroscopic system, where the relatively strong coupling
between the qubit and the environment yields appreciable decoherence to the quantum states of the qubit. The sensitivity of a qubit to
noise can often be optimized by operating the system at an optimal point (i.e., the degeneracy point)~\cite{Vion}. In the tunable phase-slip flux qubit, the optimal points are at $f=0.5+N$,
with $N$ being an integer~[see, e.g., Fig.~\ref{fig3}(a)], where the flux dispersion has
no slope and the transition frequency of the qubit cannot be changed by the first-order flux noise. Unfortunately, it is impossible to lock the qubit exactly at an optimal point because of the flux noise. In practice, the qubit will
easily leave the point and the transition frequency varies accordingly. However, similar to a transmon qubit, this effect can be much reduced in the enhance tunable phase-slip flux qubit, because
an reduction of the ratio $E_L/E_S$ leads to an exponential decrease of the flux
dispersion and consequently the transition frequency of the qubit is extremely stable with
respect to the flux noise.

In summary, a tunable phase-slip flux qubit is proposed by using a charge-SQUID that is formed by two phase-slip junctions connected in series with a superconducting island. In addition to the tunability by an externally applied flux in the qubit loop, the tunable phase-slip flux qubit is also controlled by the voltage applied on the island via a gate capacitance. Interestingly, this tunable phase-slip flux qubit is exactly dual to a tunable charge qubit controlled by both the gate voltage and the magnetic flux in the dc-SQUID. Also, we show that the inductance energy of a phase-slip flux qubit can be reduced using a large inductance inserted in the qubit loop, so that energy levels of the qubit become flattened and effects of the dominating flux noise are consequently suppressed.  As dual to a transmon qubit, this enhanced phase-slip flux qubit can exhibit improved quantum coherence.

This work was supported by the National Basic Research Program of China Grant Nos.~2011CBA00304 and 2009CB929302,
and the National Natural Science Foundation of China Grant Nos. 60836001, 91121015 and 61106121.


\end{document}